\documentclass[sigconf,screen,nonacm]{acmart}

\usepackage{macros/packages}
\usepackage{macros/macros}
\usepackage{macros/shorthands}

\begin{document}

\title{COSMIC: Enabling Full-Stack Co-Design and Optimization of Distributed Machine Learning Systems}

\settopmatter{authorsperrow=3}

\author{Aditi Raju}
\affiliation{
  \institution{Harvard University}
  \city{Cambridge}
  \state{MA}
  \country{USA}
}
\authornote{These authors contributed equally to this research.}
\email{araju@college.harvard.edu}

\author{Jared Ni}
\authornotemark[1]
\affiliation{
  \institution{Harvard University}
  \city{Cambridge}
  \state{MA}
  \country{USA}
}
\email{jaredni@college.harvard.edu}

\author{William Won}
\authornotemark[1]
\affiliation{
  \institution{Georgia Institute of Technology}
  \city{Atlanta}
  \state{GA}
  \country{USA}
}
\email{william.won@gatech.edu}

\author{Changhai Man}
\authornotemark[1]
\affiliation{
  \institution{Georgia Institute of Technology}
  \city{Atlanta}
  \state{GA}
  \country{USA}
}
\email{cman8@gatech.edu}

\author{Srivatsan Krishnan}
\authornotemark[1]
\affiliation{
  \institution{Harvard University}
  \city{Cambridge}
  \state{MA}
  \country{USA}
}
\email{srivatsan@g.harvard.edu}

\author{Srinivas Sridharan}
\affiliation{
  \institution{NVIDIA}
  \city{Santa Clara}
  \state{CA}
  \country{USA}
}
\email{srisridharan@nvidia.com}

\author{Amir Yazdanbakhsh}
\authornotemark[2]
\affiliation{
  \institution{Google DeepMind}
  \city{Mountain View}
  \state{CA}
  \country{USA}
}
\email{ayazdan@google.com}

\author{Tushar Krishna}
\authornotemark[2]
\affiliation{
  \institution{Georgia Institute of Technology}
  \city{Atlanta}
  \state{GA}
  \country{USA}
}
\email{tushar@ece.gatech.edu}

\author{Vijay Janapa Reddi}
\affiliation{
  \institution{Harvard University}
  \city{Cambridge}
  \state{MA}
  \country{USA}
}
\authornote{Equal advising.}
\email{vj@eecs.harvard.edu}

\renewcommand{\shortauthors}{A. Raju et al.}

\begin{abstract}

Large-scale machine learning models necessitate distributed systems, posing significant design challenges due to the large parameter space across distinct design stacks.
Existing studies often focus on optimizing individual system aspects in isolation.
This work challenges this limitation and introduces \ours{}, a full-stack distributed machine learning systems environment enabling end-to-end simulation and agent-based design space exploration.
To facilitate efficient exploration and optimization across the entire stack, we introduce Parameter Set Architecture\textemdash{}an abstraction concept analogous to the instruction set architecture\textemdash{}abstracting away configuration complexities of agent-based search methods.
Case studies demonstrate \ours{}'s ability to consolidate parameters across multiple layers of design abstraction, discovering eight non-obvious high-performance system configurations across four transformer-based models with up to 175 billion parameters.
By optimizing across the stack, \ours{} full-stack optimization delivers 1.50--48.41 $\times$ higher performance compared to the isolated single-stack optimization.

\end{abstract}


\maketitle

\section{Introduction}\label{sec:Introduction}

Recent advances in large-scale distributed machine learning (ML) models have revolutionized numerous applications, from protein folding~\cite{jumper2021highly} and mathematical reasoning~\cite{trinh2024solving} to performance optimization~\cite{pie_iclr_2024_spotlight,garg2022deepperf} and high-quality video generation~\cite{singer2022makeavideo,gupta2023photorealistic}.
These capabilities are enabled by increasingly sophisticated transformer-based models such as Google Gemini~\cite{google2023gemini}, OpenAI ChatGPT~\cite{openai2022chatgpt} and Sora~\cite{liu2024sora}, Claude Sonnet~\cite{anthropic2024sonnet}, and Meta LLaMA~\cite{meta2023llama}.
The scale of these models continues to grow exponentially, with parameters now reaching from billions to trillions~\cite{rajbhandari2020zero,dash2023optimizing}, and no indication of slowing down\textemdash{}their memory requirements now far exceed the tens of GBs of storage available on a single compute node~\cite{nvidia2020a100}.

\insertFigure{BackgroundParallelization}{
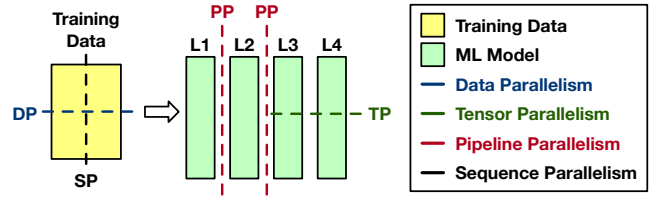
Common parallelization strategies common in distributed ML: DP, SP, PP, and TP.
}{1}{-2em}{-1em}

Executing these models poses significant resource and environmental challenges due to their unprecedented computational demands.
The scale of the problem is stark: training GPT-3 on a single NVIDIA V100 graphics processing unit~(GPU) would require an impractical 355 years~\cite{li2020singlegpu}.
Consequently, \emph{large-scale distributed ML} has become essential~\cite{shoeybi2020megatronlm,lepikhin2020gshard,singh2023ted,li2020pytorch}, requiring sophisticated strategies to partition and distribute models and data across thousands of neural processing units~(NPUs), including tensor processing units~(TPUs)~\cite{Jouppi20tpuv2,Jouppi2023tpuv4} and GPUs~\cite{nvidia2020a100,nvidia2023h100,amd2023mi300}.
The complexity of these distributed ML systems spans multiple interconnected design layers\textemdash{}\emph{Workload, Collective, Network, and Compute}\textemdash{}each incorporating critical elements such as model partitioning methods~\cite{Valiant90dp,shoeybi2020megatronlm}, synchronization schedules~\cite{rashidi22themis}, communication protocols~\cite{Thakur2005mpich}, and foundational compute technologies~\cite{Jouppi20tpuv2,Jouppi2023tpuv4}.
Given the aggregate influence of these components on ML execution performance, a co-design approach is essential.

Optimizing large-scale distributed ML systems involves three main challenges: (i)~parameterizing every component within the distributed ML stack, (ii)~creating comprehensive simulation capabilities to predict the cost of executing multi-node deep learning~(DL) models, and (iii)~productively and efficiently exploring the extensive parameter space for customization.
The scale of this co-design space is staggering\textemdash{}even with just four parallelization dimensions (DP, SP, TP, PP), each ranging from 1 to 1024, there are 286 possible combinations. This complexity is further compounded by additional parameters such as link bandwidth allocation, chunks per collective, and multi-dimensional collective optimization~\cite{cho2019blueconnect}.
Furthermore, combined with collective scheduling options, choices for communication algorithms, topology options, and possible NPU distributions, the design space explodes to approximately $7.69 \times 10^{13}$ potential configurations~(see \autoref{tab:schema}).
Even with an optimistic simulation time of one second per design point, an exhaustive search would require an impractical $2.44 \times 10^{6}$ years, making traditional search methods infeasible as systems scale towards larger models, higher NPU counts, and more complex topologies.

Prior design space exploration~(DSE) methods have attempted to manage this complexity by focusing on isolated components of the system stack~\cite{lu2017flexflow,lin2024uniap,won2024tacos,Shah2023taccl,won2024libra,zhao2024efficient,kao2020gamma,wang2023topoopt}.
Although some approaches have employed agent-based methods to optimize parameters within specific layers~\cite{lu2017flexflow,kao2020gamma}, this isolated optimization strategy can lead to several limitations.
First, optimizing individual components in isolation risks overfitting to local optima, potentially missing critical interactions between different layers of the stack.
Second, the lack of a unified abstraction framework makes it challenging for domain experts to effectively utilize these optimization methods across different system configurations.
Third, focusing on subset optimizations limits the agent's ability to discover non-obvious solutions that might emerge from cross-layer interactions.
These limitations become particularly acute as model sizes and system complexity continue to grow, necessitating a more comprehensive approach to system optimization.

To address these challenges, we propose a two-part solution: (i)~Parameter Set Architecture~(\psa), an abstraction layer for defining and managing the complex parameter space of distributed ML systems, and (ii)~\ours\footnote{\textbf{C}ross-stack \textbf{O}ptimization and \textbf{S}ystem \textbf{M}odeling with \textbf{I}ntelligent \textbf{C}o-design}, a framework that leverages \psa\ to enable efficient artificial intelligence~(AI) agent-based exploration of full-stack distributed ML platform designs. 

The \psa\ is analogous to how an instruction set architecture~(ISA) defines the interface between software and hardware.
It defines the interaction between search agents and the underlying system using a standard schema-based approach, enabling domain experts to specify searchable parameters, valid ranges, and design constraints without requiring direct agent configuration.
The \psa\ scheduler \emph{automatically} establishes the abstraction layer between agents and the design space, facilitating efficient exploration of configurations across all system layers.

Leveraging \psa, we introduce \ours, a framework that conjoins full-stack large-scale distributed ML with agent-based search capabilities.
\ours\ combines ASTRA-sim~\cite{won2023astrasim2}'s detailed distributed ML simulation capabilities with ArchGym~\cite{archgym}'s flexible ML-agent framework, creating a comprehensive environment for exploring distributed ML system designs.
By leveraging \psa's abstraction layer, \ours\ enables automatic configuration of both the simulation environment and search agents, allowing efficient exploration of designs spanning from workload parallelization strategies to network topology choices.
The framework supports multiple ML-based search algorithms\textemdash{}including Bayesian optimization~(BO), ant colony optimization~(ACO), and genetic algorithms~(GA)\textemdash{}each bringing unique exploration strategies to navigate the vast design space.
Using this approach in our evaluation, we discover several
distinct large-scale distributed system configurations, each exhibiting similar performance characteristics while offering different trade-offs in system design.
This diversity of high-performing configurations demonstrates the effectiveness of our approach in navigating the complex co-design space and identifying non-obvious solutions that would be difficult to discover through traditional optimization approaches.

This work makes the following contributions:
\begin{itemize}[leftmargin=*]
    \item We introduce Parameter Set Architecture~(\psa)\textemdash{}akin to the role of ISA in delineating the interface between software and hardware\textemdash{}\psa\ \emph{delineates the interface between search agents and the underlying system}.
    It offers a schema for defining searchable parameters and their ranges along with design constraints.
    This approach shields domain experts from the complexities of agent setup and configuration while enabling seamless exploration of design spaces comprising millions of parameters.
    
    \item We introduce \ours---a full-stack large-scale distributed ML simulation and agent-based search mechanism combined, built upon \psa.
    \ours\ can identify distinct system configurations, including workload parallelization strategies, collective algorithms, and network topologies, with similar performance traits, providing system designers with multiple viable implementation options.
    
    \item We evaluate four transformer-based models (up to 175B parameters) and perform more than six million steps across four search agents.
    Our results demonstrate that diversifying parameters across the entire design stack of distributed ML systems outperforms approaches that merely increase the design space of isolated components.
    Specifically, we show that full-stack optimization enabled by \ours{} delivers 1.50--48.41 $\times$ higher performance compared to the isolated single-stack optimization.
\end{itemize}

\section{Background}\label{sec:Background}

Distributed ML of large models involves complex interactions across multiple stacks.
To understand the design space and optimization challenges addressed by our work, we first review four key aspects of distributed ML systems: (i)~parallelization strategies that enable model and data distribution, (ii)~collective communication patterns that facilitate data synchronization, (iii)~network topologies that affect communication performance, and (iv)~compute devices that execute the ML operations.
Each of these components presents unique trade-offs and constraints that must be considered in full-stack co-design optimization.

\subsection{Parallelization Strategies}\label{subsec:Subsec2-1}

We describe four key workload parallelization approaches in distributed ML, which are summarized in~\autoref{fig:BackgroundParallelization}.

\niparagraph{Data Parallelism~(DP)}~divides the input data set into smaller batches and processes them concurrently across multiple NPUs, increasing the throughput~\cite{Valiant90dp}.
However, DP does not reduce the memory footprint requirement of each compute device.

\niparagraph{Sequence Parallelism~(SP)}~splits input activation across the sequence dimension, particularly for large language models~(LLMs)~\cite{nvidia2024sp}.
SP can yield better throughput for operations like dropout or normalization. However, SP does not reduce NPU memory footprint.

\niparagraph{Tensor Parallelism~(TP)}~vertically splits the deep neural network~(DNN) model, distributing each layer across multiple NPUs~\cite{shoeybi2020megatronlm}.
TP can help reduce the memory requirement of each NPU, but it introduces frequent and costly communication overhead.

\niparagraph{Pipeline Parallelism~(PP)}~distributes the DNN across the dimension of the ML layer~\cite{huang2019gpipe, harlap2018pipedream}.
Each NPU processes a subset of layers and forwards the activation.
PP reduces memory usage per NPU, but efficient pipeline orchestration is necessary to prevent bottlenecks.

\insertFigure{CollectiveComms}{
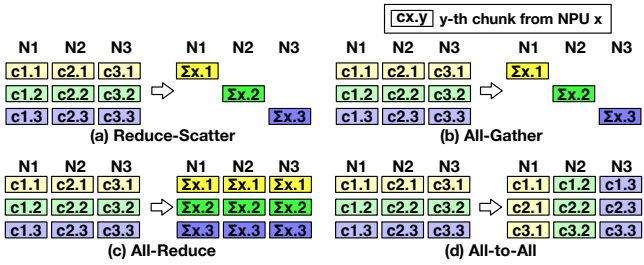
Common collective communication patterns incurred in distributed ML for NPU synchronization.}{1}{-2em}{-1em}

\subsection{Collective Communications}\label{subsec:Subsec2-2}

Due to the distributed nature of models, NPUs must periodically synchronize data, such as activations or weight gradients.
This process relies on \emph{collective communications}, specialized synchronization patterns that manage data flow across the network, processed in fine-grained \textit{chunks}~\cite{Thakur2005mpich}.
\autoref{fig:CollectiveComms} summarizes commonly used collective patterns in distributed ML execution.

\niparagraph{\reducescatter}~reduces corresponding chunks from every NPU and then distributes the result.

\niparagraph{\allgather}~broadcasts a chunk of each NPU to all other NPUs. This makes all chunks accessible to the entire cluster.

\niparagraph{\allreduce}~is the most commonly used collective~\cite{klenk2020innetwork}.
All chunks are reduced and broadcasted, which is essential for aggregating weight gradients to synchronize ML models.

\niparagraph{\alltoall}~patterns occur when each NPU generates and transfers dedicated chunks for all other NPUs, such as gating functions in Mixture-of-Experts~(MoE) models~\cite{shazeer2017outrageously}.

Several \emph{collective communication algorithms} define the traffic patterns to implement and execute collective communication patterns.
Collective communication libraries~(CCLs), such as NCCL~\cite{nvidia2024nccl}, implement multiple collective algorithms, including \ringAlg~(RI)~\cite{chan2006ring}, \direct~(DI)~\cite{rashidi2020astrasim}, \rehd~(RHD)~\cite{Thakur2005mpich}, and \dbt~(DBT)~\cite{Jeaugey2019dbt}.

\subsection{Network Topology}\label{subsec:Subsec2-3}

The network fabric processes message transfers in distributed DL systems.
Prior work in network technology, such as high-speed interconnects like InfiniBand~\cite{mellanox2003infiniband} improves performance and efficiency.
Different network technologies have distinct performance and cost implications, leading to multiple network fabrics in state-of-the-art clusters.
For example, Google Cloud TPU~\cite{Jouppi2023tpuv4} uses a 3D~Torus topology with electrical and photonic networks, whereas NVIDIA uses multi-dimensional switches through NVLink~\cite{nvidia2024nvlink}, NVSwitch~\cite{nvidia2024nvlink}, and InfiniBand~\cite{mellanox2003infiniband}.
Cerebras~\cite{cerebras2021cs2} and Tesla~\cite{tesla2021dojo} clusters use wafer-scale engines with on-chip networks and electrical scale-out capabilities.
We abstract these core topologies using the multi-dimensional network representation from~\cite{won2023astrasim2}.
\ringTopo~(RI), \switch~(SW), and \fc\ (FC), the core network building blocks in this notation, are summarized in \autoref{fig:TopologyBlock}.
Multi-dimensional network fabrics are represented by stacking these blocks, with link bandwidth and latency information for each dimension (e.g., 3D~Torus can be captured as [RI, RI, RI]).

\insertFigure{TopologyBlock}{
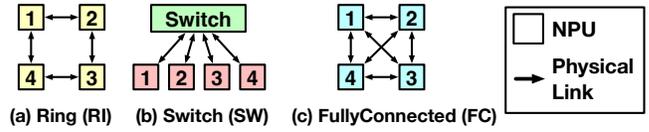
Network topology building blocks we considered to construct multi-dimensional topologies.
}{1}{-2em}{-1em}

\subsection{Compute Devices}

Compute devices refer to the components that run all compute operators of ML workloads.
GPUs are the most prevalent compute accelerators in ML systems, with state-of-the-art examples including A100~\cite{nvidia2020a100} and H100~\cite{nvidia2023h100} from NVIDIA, and MI300~\cite{amd2023mi300} from AMD.
For additional energy efficiency, there also exist custom accelerators such as TPU~\cite{Jouppi20tpuv2} from Google, Trainium~\cite{amazon2020trainium} from Amazon, CS2~\cite{cerebras2021cs2} from Cerberas.
In this work, we refer to the compute devices as NPUs.

In our case studies, we model the compute device by three parameters: \textit{peak-perf}, \textit{local-mem-bw}, and \textit{memory-capacity}.
With the first two parameters, we can create a simple roofline model to get the runtime of each operator. 
With the last parameter, we can apply the constraint of the memory capacity, which will affect the design space for parallelization strategies.

\section{Motivation for Full-Stack Co-Design}\label{sec:System}

\insertFigure{graphs_cycles_2x2}{
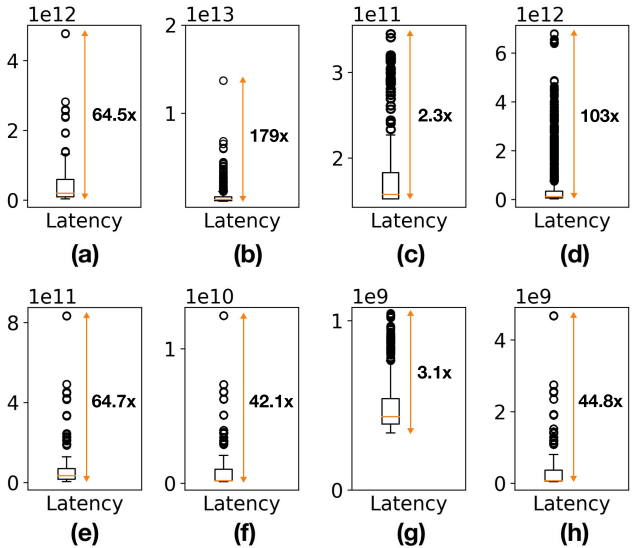
(a)~shows latency spread for training GPT3-175B just varying the workload parameters (i.e., workload-only search for System 2, see~\autoref{subsec:experiments}).
Notably, the parallelization optimal for the target cluster achieved 64.5$\times$ better performance compared to the worst case, highlighting co-optimization potential.
(b)--(d) shows workload+network, workload+collective, and full-stack optimization results for \gptlarge{}.
(e)~latency spread for workload-only DSE for \gptsmall{}, (f)~workload-only DSE for \vitlarge{}, (g)~full-stack DSE for \vitlarge{}, and (h)~full-stack DSE for \vitsmall{}.
\label{fig:spread_cycles}
}{1}{-2em}{-1.5em}

The rapid growth of LLMs requires efficient distributed ML systems.
Optimizing these systems involves a complex co-design space that spans workload, system, network, and compute characteristics.

\subsection{Opportunities for Full-stack Optimization}

Traditional parameter tuning approaches in distributed ML systems often focus on optimizing individual components in isolation~\cite{lu2017flexflow,kao2020gamma,wang2023topoopt}.
This isolated optimization fails to capture the complex interactions between different layers of the system stack, potentially missing significant performance opportunities. 

Our preliminary analysis, summarized in~\autoref{fig:spread_cycles}, demonstrates the importance of considering these cross-layer interactions.
For instance, \autoref{fig:spread_cycles}(a) fixes the physical cluster and observes how the workload parallelization strategy can impact the GPT3-175B training latency.
Even when considering only workload-level parameters, the latency can vary up to 64.5$\times$.

This significant improvement from tuning a single design stack suggests substantial untapped potential in the broader design space.
When we expand our search to simultaneously optimize the entire design stack, the performance improvements amplify dramatically.
As shown in~\autoref{fig:spread_cycles}(d), the co-designed cluster can achieve latency reductions of up to 103$\times$.
This multiplicative effect reveals the presence of powerful cross-layer optimization opportunities.

These results highlight two insights: (i)~neglecting parameter optimization across layers can potentially lead to significant performance penalties in distributed ML systems. Local optima at individual layers may conflict with global system efficiency. Also, (ii)~the vast co-design space with its intricate dependencies necessitates intelligent search agents capable of efficiently identifying better configurations across workload, system, and network components.

\subsection{Challenges of Tackling the Co-design Space}\label{subsec:challenge_codesign}

While full-stack optimization is important in efficient distributed ML systems, their design space suffers from a combinatorial explosion, making exhaustive search impractical.
Consider a system with a 4D network topology and 1,024 NPUs.
Different design knobs and a number of potential design points are summarized in~\autoref{tab:schema}.

Parallelization dimensions (DP, PP, SP, TP), each ranging between $(1, 1, 1, 1024)$, $(1, 1, 2, 512)$, $\cdots$, $(1024, 1, 1, 1)$, already creates 286 potential options.
This can be multiplied with two collective scheduling policy choices, 256~(=$4^4$) 4D collective communication algorithms, 81~(=$3^4$) physical network topology options, etc., further exploding the design space.
The total combination of the co-design space for this 1,024~NPU cluster results in about $7.69 \times 10^{13}$ potential designs.
Even with an optimistic simulation time of one second per design point, an exhaustive search would take an impractical $2.44 \times 10^{6}$ years.
As systems scale towards larger models, higher NPU counts, and more complex topologies, this combinatorial explosion renders naive search methods infeasible.

\subsection{Toward an ML-Driven Approach}

The exponential growth of distributed ML platform design spaces has led to the increasing adoption of ML-driven search agents for system optimization.
These agents\textemdash{}from Bayesian optimization~(BO)~\cite{Kushner1964bo,movckus1975bo} and genetic algorithms~(GA)~\cite{katoch2021ga} to random walk~(RW) exploration~\cite{pearson1905rw}\textemdash{}have demonstrated remarkable success in navigating high-dimensional parameter spaces~\cite{kao2020gamma,Esmaeilzadeh2024mlsearch}.
Their diverse strengths complement different aspects of the search problem: BO efficiently finds global optima with limited samples, GA excels at discrete parameter spaces, and even simple RW approaches provide valuable baseline exploration capabilities.

However, leveraging these ML agents effectively creates a significant knowledge barrier.
Current optimization frameworks like \archgym~\cite{archgym} force an unnecessary coupling between domain expertise and ML knowledge: system designers must understand the internals of the ML agent to define observation and action spaces, while ML experts need deep domain knowledge to properly configure the search space.
This coupling becomes problematic when parameters require modification across different layers of the distributed ML platform stack, when new design spaces or system configurations need to be explored, or when different ML agents need to be evaluated on the same design space.
This intricate interdependency between system design knowledge and ML expertise creates a bottleneck in the optimization process, highlighting the need for a systematic decoupling approach.

\begin{table}[t]
\centering

\caption{
\psa{} schema to capture distributed ML design space.
The number of design points are retrieved by assuming a 4D network with 1,024 NPUs (see~\autoref{subsec:challenge_codesign}).
}
\label{tab:schema}
\vspace{-1em}
\resizebox{\linewidth}{!}{

\begin{tabular}{llr}
\toprule
\multicolumn{1}{l|}{\textbf{Workload Knob}} & \multicolumn{1}{l|}{\textbf{Value Range}} & \multicolumn{1}{l}{\textbf{\#Points}} \\
\midrule
\multicolumn{1}{l|}{DP} & \multicolumn{1}{l|}{\{1, 2, 4, 8, $\cdots$, 512, 1024\}} & . \\
\multicolumn{1}{l|}{PP} & \multicolumn{1}{l|}{\{1, 2, 4, 8, $\cdots$, 512, 1024\}} & . \\
\multicolumn{1}{l|}{SP} & \multicolumn{1}{l|}{\{1, 2, 4, 8, $\cdots$, 512, 1024\}} & 286 \\
\multicolumn{1}{l|}{Weight Sharded} & \multicolumn{1}{l|}{\{0, 1\}} & 2 \\ \hline
\multicolumn{1}{l|}{\textbf{Collective Knob}} & \multicolumn{1}{l|}{\textbf{Value Range}} & \textbf{\#Points} \\ \hline
\multicolumn{1}{l|}{Scheduling Policy} & \multicolumn{1}{l|}{\{LIFO, FIFO\}} & 2 \\
\multicolumn{1}{l|}{Collective Algorithm} & \multicolumn{1}{l|}{MultiDim \{Ring, Direct, RHD, DBT\}} & 256 \\
\multicolumn{1}{l|}{Chunks per Collective} & \multicolumn{1}{l|}{\{1, 2, 3, 4, $\cdots$, 32\}} & 32 \\
\multicolumn{1}{l|}{Multi-dim Collective} & \multicolumn{1}{l|}{\{Baseline, BlueConnect\}} & 2 \\ \hline
\multicolumn{1}{l|}{\textbf{Network Knob}} & \multicolumn{1}{l|}{\textbf{Value Range}} & \textbf{\#Points} \\ \hline
\multicolumn{1}{l|}{Topology} & \multicolumn{1}{l|}{MultiDim \{Ring, Switch, FC\}} & 81 \\
\multicolumn{1}{l|}{NPUs per Dim} & \multicolumn{1}{l|}{MultiDim \{4, 8, 16\}} & 81 \\
\multicolumn{1}{l|}{Bandwidth per Dim} & \multicolumn{1}{l|}{MultiDim \{100, 200, 300, 400, 500\}} & 625 \\ \hline
\multicolumn{2}{l|}{\textbf{Total \#Points}} & \multicolumn{1}{l}{\textbf{$7.69 \times 10^{13}$}} \\ \hline
\multicolumn{3}{l}{\textbf{Constraints}} \\ \hline
\multicolumn{3}{l}{product (DP, SP, PP) $\le$ (Number of NPUs) = 1,024} \\
\multicolumn{3}{l}{product (NPUs per Dim) = (Number of NPUs) = 1,024} \\
\bottomrule
\end{tabular}

}
\vspace{-1em}
\end{table}

\insertFigureWide{PSA_Architecture}{
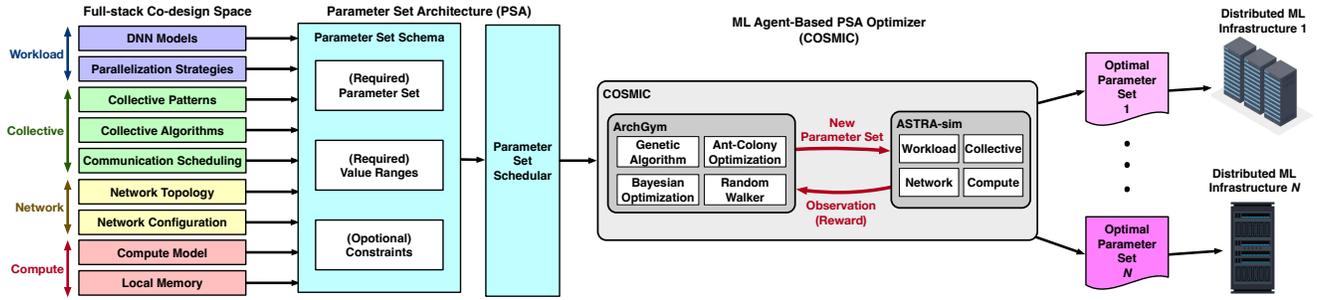
Summary of (i)~\psa\ to capture the full-stack distributed ML design space and (ii)~ML-based \psa\ optimization framework~(\ours) to design new distributed ML infrastructures.
\label{fig:PSA_architecture}
}{1}{-2em}{-0em}

\section{\ours: Parameter-Based Approach for DSE}\label{sec:Cosmic}

To address the myriad challenges, we introduce \ours, a framework that implements Parameter Set Architecture (\psa) to enable systematic exploration of distributed ML design spaces. \ours\ creates a clean separation between design parameters and search mechanics through two key capabilities: (i)~a structured parameter definition schema and (ii)~an automated scheduler that configures ML agents. This approach eliminates the need for domain experts to understand the internals of ML agents while allowing the ML agents to efficiently explore the hardware design spaces.

\subsection{Design Principles}

\ours's design is guided by three fundamental principles that together enable efficient exploration of distributed ML design spaces:

\niparagraph{Separation of Concerns.}
The primary goal of \ours\ is to establish clear boundaries between domain knowledge and ML expertise.
Domain experts should be able to express complex design spaces and constraints using familiar terminology and concepts, without having to understand the internals of the ML agent.
Conversely, ML agents should be able to efficiently explore these spaces without requiring deep knowledge of system design.
This separation enables each group to focus on their core competencies while still working or collaborating effectively together.

\niparagraph{Automated Configuration.}
\ours\ eliminates manual agent configuration through automation.
Rather than requiring experts to manually define observation spaces, action spaces, and reward functions, \ours\ translates domain-specific parameter definitions into ML-agent configurations.
This automation not only reduces the potential for errors but also enables rapid experimentation with different ML agents on the same design space, as the configuration process is handled systematically rather than manually.

\niparagraph{Flexible Expression.}
The framework must accommodate the complexity of distributed ML systems while maintaining usability.
This means supporting simple single-stack optimizations (like parallelization strategies) and complex multistack design spaces (involving workload, collective, and network parameters).
Moreover, we must capture real-world constraints and parameter dependencies without sacrificing the clarity of the parameter specification process.
This flexibility ensures that \ours\ can address both current optimization needs and future expansions of the design space.

These principles work together to create an abstraction layer that makes full-stack optimization more accessible while maintaining the power and flexibility needed for complex distributed ML systems, whose schema we discuss next.

\subsection{Parameter Set Schema}

\ours\ relies on its schema definition, which provides a systematic way to express and constrain the design space.
The schema serves as a contract between domain experts and ML agents, ensuring that all explored configurations are valid and meaningful.
To achieve this, the schema is structured around three components:

\niparagraph{Parameter Set.}
This component defines the specific parameters to be explored, encompassing both single-layer and multi-stack design spaces.
These parameters span workload characteristics (e.g., parallelization strategies), collective configurations (e.g., collective algorithms), and network specifications.
Each parameter represents a distinct dimension in the design space that can be tuned to optimize the system performance.

\niparagraph{Value Range.}
To ensure practical exploration, each parameter is associated with a defined target range of valid values, carefully selected by experts in the domain.
These ranges serve dual purposes: (i)~they control the \ours\ scheduler's configuration space, and (ii)~delimit the search boundaries for ML agents.
This structured approach prevents agents from exploring invalid or impractical configurations while maintaining flexibility within meaningful bounds.

\niparagraph{Constraints.}
Real-world systems rarely have independent parameters.
The schema, therefore, allows domain experts to specify cross-parameter dependencies using additional constraints.
These constraints capture relationships between parameters, ensuring that the generated configurations respect the requirements at the system level and technological limitations.
Although this increases complexity, it is crucial to generate viable system configurations.

This schema definition allows for flexible and expressive definitions of the design space, capturing both individual parameter ranges and inter-parameter dependencies.
The workload stack captures the number of target NPUs in a system and their potential parallelization strategies.
The collective stack captures collective-level communication configurations, such as collective scheduling policy, collective algorithms, or whether to enable BlueConnect~\cite{cho2019blueconnect} multi-dimensional optimization.
The network stack encapsulates the information about the topology and associated network bandwidth.
Cross-stack correlation information is also being captured through constraints.
For example, the parallelization strategy is limited by the number of physical NPUs in the cluster.

\subsection{Parameter Set Scheduler}

Parameter Set Scheduler~(PSS) is there for automating the configuration of both search environments and agents.
Adapting search mechanisms to changes in parameter sets traditionally requires manual adjustments to both the environment and the agent's action space.
This manual configuration exposes the inherent complexity of heavy domain knowledge, hindering efficient exploration.
The PSS overcomes this challenge by abstracting away this complexity, enabling seamless integration and execution of search agents for design exploration. Below we describe the main design components.

\niparagraph{Environment Side Configuration.}
The search environment encapsulates the core cost model (e.g., simulators).
It receives design parameters as input and estimates desired performance metrics like latency or communication overhead.
For efficient interaction with search agents, the environment must clearly define (i)~action space, (ii)~observation space, and (iii)~constraints to prevent ineffectual simulations with invalid parameter combinations.
The PSS automates this process by defining the action and observation spaces and incorporating constraints based on the parameter set schema. 

\niparagraph{Agent Side Configuration.}
The PSS configures various aspects of the search agent to ensure efficient DSE.
For effective interaction with the environment interface, the agent must clearly define: (i)~the agent's search space, (ii)~the type of each parameter (integer, categorical, boolean, float, etc.), (iii)~the step size for each parameter as defined in the parameter schema, and (iv)~the configuration of reward function parameters.
The PSS automates this process, allowing search agents to effectively navigate the design space with well-defined parameters and rewards.

\subsection{\ours\ Implementation}\label{sec:Methodology}

To prove and show the effectiveness of our parameter-centric approach, we implement \ours, a framework that enables ML-driven full-stack optimization of distributed ML systems.
\ours\ integrates two key components: \astrasim{}'s detailed, end-to-end distributed ML simulation capabilities, and \archgym{}'s flexible ML-agent search mechanism.
We describe how our parameter-centric approach enables this integration and discuss the key components.

\astrasim{} is the simulation engine for \ours.
As an open-source, full-stack distributed ML training simulator, it models intricate interactions across workload characteristics, system parameters, network topology, and compute resources.
 \astrasim{} has been cross-validated against real system measurements, ensuring accurate performance analysis of distributed ML.

\archgym{} serves as the ML exploration backbone of \ours.
It streamlines ML-driven architecture exploration by providing a standardized interface between search agents and architectural simulators through the OpenAI Gym library.
This standardization eliminates the need for manual agent configuration, allowing pre-configured ML agents to efficiently explore parameter spaces.

Our parameter-centric abstraction enables seamless integration of these components through a well-defined agent-environment interaction loop.
At the heart of this integration is the PSS, which translates domain-specific parameter definitions into action spaces that ML agents can understand and explore.
When an agent selects a particular system configuration, it submits this action to the \astrasim{} simulation backend.
 \astrasim{} then performs a detailed simulation of the distributed ML system with the specified parameters, generating performance metrics and corresponding reward signals.
These results are processed by \archgym{}'s environment layer, which uses this feedback to guide agents toward increasingly better configurations.
This continuous feedback loop, mediated by our parameter-centric abstraction, enables efficient exploration of the design space while maintaining separation between domain expertise and ML implementation details.
This \psa-enabled integration significantly simplifies the exploration of full-stack distributed ML design spaces by abstracting away the complexity of component interaction while maintaining the flexibility to evaluate diverse system configurations.

\astrasim{} receives a workload trace to be simulated.
Distinct workload parallelization yields a completely different workload trace.
Consequently, to enable the exploration of workload knobs as defined in the \psa, we equipped the \ours\ framework with a Workload Trace Generator~(WTG).
The WTG defines multiple trace templates, which capture and represent the ML workload architecture in terms of atomic operators.
For example, these templates show structures like multi-head attention mechanisms.
Compared to the complete compute trace, such a template maintains flexibility as (i)~it is not represented in exact numbers and instead uses numeric symbols, e.g., \{B, S, D, H\}, and (ii)~the partitioning of tensors is also symbolically captured using the workload knobs, e.g., \{tp, dp\}.
When the \psa\ search knobs are provided by the PSS, the WTG translates the trace template into an actual trace to be simulated by \astrasim{}.
To retrieve the trace, it first substitutes the numeric symbols in the template using the actual \psa\ knobs to get actual tensor shapes and sizes.
Then, the generator analyzes all tensors, especially their shape and producer/consumer NPU information, and injects collective communications whenever necessary.

\section{Experimental Setup}\label{subsec:experiments}

This section outlines the target workloads, baseline system configurations, and experimental setup used to demonstrate the advantages of \psa\ and \ours.

\subsection{Target Workloads and Systems}

\begin{table}[t]
\centering

\caption{Target Workloads.}
\label{table:typical_workloads}
\vspace{-1em}
*: To expedite runtime, we simulate 4 layers per model and scale memory capacities to correctly reflect memory constraints. In post-processing, we re-scale latency and memory to reflect the whole model performance. 
\resizebox{\linewidth}{!}{
\begin{tabular}{c|cccc}
\toprule
\textbf{Parameters} & GPT3-175B & GPT3-13B & ViT-Base & ViT-Large\\ \midrule
\begin{tabular}[c]{@{}c@{}}\textbf{Number Layers*}\end{tabular} & 96 & 40 & 12 & 24\\\hline
\begin{tabular}[c]{@{}c@{}}\textbf{Embedding Dimension}\end{tabular} & 12288 & 5140 & 768 & 1024 \\\hline
\begin{tabular}[c]{@{}c@{}}\textbf{FFN Dimension}\end{tabular} & 49152 & 20560 & 3072 & 4096\\\hline
\begin{tabular}[c]{@{}c@{}}\textbf{Sequence Length}\end{tabular} & 2048 & 2048 & 256 & 256 \\\hline
\begin{tabular}[c]{@{}c@{}}\textbf{Number Heads}\end{tabular} & 96 & 40 & 12 & 16

\\ \bottomrule
\end{tabular}}

\end{table}

\begin{table}[t]
\centering

\caption{System configurations used for evaluation.}
\label{table:system_setup}
\vspace{-1em}

\resizebox{\linewidth}{!}{

\begin{tabular}{lccc}
\toprule
\textbf{Knobs} & \textbf{System 1} & \textbf{System 2} & \textbf{System 3} \\ \midrule
\multicolumn{4}{l}{\textbf{Collective Knob}} \\ \midrule
Collective Algorithm & {[}RI, RI, RI, RHD{]} & {[}RI, DI, RI, RHD{]} & {[}DI, RHD, RI, RI{]} \\ \midrule
\multicolumn{4}{l}{\textbf{Network Knob}} \\ \midrule
Topology & {[}RI, RI, RI, SW{]} & {[}RI, FC, RI, SW{]} & {[}FC, SW, RI, RI{]} \\
NPUs per Dim & {[}4, 4, 4, 8{]} & {[}4, 8, 4, 8{]} & {[}8, 16, 4, 4{]} \\
Bandwidth per Dim & {[}200, 200, 200, 50{]} & {[}375, 175, 150, 100{]} & {[}900, 100, 50, 12.5{]} \\ \midrule
\multicolumn{4}{l}{\textbf{Compute Knob}} \\ \midrule
Compute Performance & 459 & 10 & 900 \\
Local Mem BW & 2765 & 50 & 3000 \\
\bottomrule
\end{tabular}

}

\vspace{-1em}
\end{table}

We leverage four LLMs of varying scales to evaluate our approach: \gptsmall, \gptlarge~\cite{brown2020gpt}, \vitsmall, and \vitlarge~\cite{dosovitskiy2021vit}.
Details of each model are summarized in~\autoref{table:typical_workloads}.

For baseline targets, we evaluate three systems differing in the number of NPUs and corresponding real-world configurations:
\begin{itemize}[leftmargin=*]
    \item \textbf{System 1}: Represents a cluster of 512 Google TPUv5p devices~\cite{google2023tpuv5p}.
    \item \textbf{System 2}: Based on a 4D network topology comprising 1,024 NPUs, as described in~\cite{rashidi22themis}.
    \item \textbf{System 3}: Serves as a proxy for a 2,048-NPU system using NVIDIA H100 GPUs~\cite{nvidia2023h100}.
\end{itemize}

Detailed specifications for each target system are provided in~\autoref{table:system_setup}.

\begin{footnotesize}
\begin{table}[t]
\centering

\caption{Target \psa\ used for evaluations in this work.}
\label{table:target_psa}
\vspace{-1em}

\begin{tabular}{ll}
\toprule
\textbf{Workload Knob} & \textbf{Value Range} \\
\midrule
DP & \{1, 2, 4, 8, $\cdots$, 1024, 2048\} \\
PP & \{1, 2, 4\} \\
SP & \{1, 2, 4, 8, $\cdots$, 1024, 2048\} \\
Weight Sharded & \{0, 1\} \\
\midrule
\textbf{Collective Knob} & \textbf{Value Range} \\
\midrule
Scheduling Policy & \{LIFO, FIFO\} \\
Collective Algorithm & MultiDim \{Ring, Direct, RHD, DBT\} \\
Chunks per Collective & \{2, 4, 8, 16\} \\
Multi-dim Collective & \{Baseline, BlueConnect\} \\
\midrule
\textbf{Network Knob} & \textit{\textbf{Value Range}} \\
\midrule
Topology & MultiDim \{Ring, Switch, FC\} \\
NPUs per Dim & MultiDim \{4, 8, 16\} \\
Bandwidth per Dim & MultiDim \{50, 100, 150, $\cdots$, 450, 500\} \\
\midrule
\multicolumn{2}{l}{\textbf{Constraints}} \\
\midrule
\multicolumn{2}{l}{product (DP, SP, PP) $\le$ (Number of NPUs)} \\
\multicolumn{2}{l}{product (NPUs per Dim) = (Number of NPUs)} \\
\bottomrule
\end{tabular}

\vspace{-1em}
\end{table}
\end{footnotesize}

\subsection{\psa\ Parameters}

Each experiment selects a specific target system and employs \ours, guided by \psa, to optimize either a fixed stack for baseline comparisons or a full-stack configuration through automated search.
As discussed in~\autoref{sec:Cosmic}, the optimization performed by \ours\ is enabled by the abstraction layer provided by \psa.
\autoref{table:target_psa} summarizes the \psa\ configurations used in our evaluation, specifying the target values for each distributed ML design parameter exposed by \psa\ during system optimization.

\subsection{ML Search Agents}\label{subsec:Subsec5-1}

During ML-based optimization, \ours\ utilized one of four ML agents.
Thanks to the abstraction provided by \psa, the implementation of distributed ML modeling is entirely decoupled from agent-based search.
As a result, any reinforcement learning~(RL) algorithm can be integrated with \ours\footnote{
We did not evaluate RL methods in this work due to their significantly higher simulation time per sample (approximately 20--30 minutes).
}.

\niparagraph{Random Walker (RW)}~\cite{pearson1905rw}.
For RW, we vary the population size, i.e., the number of agents.

\niparagraph{Genetic Algorithm (GA)}~\cite{katoch2021ga}.
For GA, we tune both the population size and the mutation probability.

\niparagraph{Ant Colony Optimization (ACO)}~\cite{dorigo1999aco}.
For ACO, we adjust the number of ants (agents), the greediness factor, and the pheromone evaporation rate.

\niparagraph{Bayesian Optimization (BO)}~\cite{movckus1975bo}.
For BO, we randomize the surrogate model by varying the random seed of the underlying Gaussian process.

\subsection{Optimization Objectives}\label{subsec:experiments:Objectives}

\ours\ is designed to maximize ML performance, specifically by minimizing total ML runtime.
However, unconstrained exploration of the network configuration space may lead \ours\ to select overly resource-intensive setups---for example, assigning the maximum available bandwidth to each network dimension.
To avoid this, we introduce additional constraints to regulate the network search space.

\niparagraph{Runtime per BW/NPU.}
In this setup, \ours\ aims to maximize ML performance while limiting the bandwidth allocated per NPU.
To achieve this, \ours\ minimizes the product of ML execution latency and the total network bandwidth consumed per NPU.
A minus-one offset is applied to prevent divide-by-zero errors in cases where invalid configurations produce a latency or bandwidth of zero.

\begin{equation*}
\text{reward}_{\text{(Perf per BW/NPU)}} = \frac{1}{\sqrt{(\text{Latency} \times \sum\text{(BW per Dim)} - 1)^{2}}}
\end{equation*}

\niparagraph{Runtime per Network Cost.}
As an alternative regulation metric, we assess ML performance in relation to network dollar cost.
We adopt the cost model proposed in~\cite{won2024libra}, and define the reward function as:

\begin{equation*}
\text{reward}_{\text{(Perf per Network Cost)}} = \frac{1}{\sqrt{(\text{Latency} \times \text{Network Cost} - 1)^{2}}}
\end{equation*}

Additionally, any parallelization strategy resulting in a memory footprint exceeding 24~GB per NPU is considered invalid and discarded.

\section{Results}

This section presents experimental results and evaluates the impact of co-optimizing different design stacks in distributed ML systems. We also examine the effect of using different ML agents.

\subsection{Full-Stack Optimizations}\label{section:Results:FullStack}

We first demonstrate the benefits of full-stack optimization using \ours. Such end-to-end optimization is made possible by the abstraction provided by \psa, as outlined in~\autoref{table:target_psa}, which enables seamless integration with ML-based DSE frameworks.

We employed \ours\ to perform DSE to optimize the training of the \gptlarge\ workload. Specifically, \ours\ tunes the \psa\ parameters to maximize the reward functions defined in~\autoref{subsec:experiments:Objectives}.

To assess the contributions of each layer, we restricted \ours\ to operate on individual stacks:
\begin{itemize}[leftmargin=*]
    \item \textbf{Workload-only search:} Optimizes parallelization parameters (DP, PP, SP) and weight sharding flags, analogous to approaches like FlexFlow~\cite{lu2017flexflow} and UniAP~\cite{lin2024uniap}.
    \item \textbf{Collective-only search:} Tunes collective algorithms, serving as a proxy for communication-optimization works.
    \item \textbf{Network-only search:} Adjusts network topology and resource allocation, reflecting system-level optimization efforts.
    \item \textbf{Full-stack search:} Optimizes all parameters exposed by \psa.
\end{itemize}

\ours\ was evaluated on System 1 (512 NPUs) and System 2 (1,024 NPUs). Results are shown in~\autoref{fig:ResultPerfBW} and~\autoref{fig:ResultPerfCost}.

\insertFigure{ResultPerfBW}{
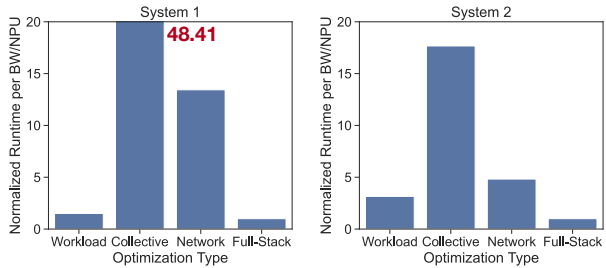
ML runtime per BW/NPU for the \gptlarge\ workload. 
The base systems were System 1 and System 2. 
Each bar represents a scenario where \ours\ is restricted to workload-only, collective-only, network-only, or full-stack optimization.
Results are normalized to the full-stack outcome.
}{1}{-2.5em}{0em}

\insertFigure{ResultPerfCost}{
Normalized ML runtime per network dollar cost for \gptlarge. 
Similar to~\autoref{fig:ResultPerfBW}, values are normalized to the corresponding full-stack search result.
}{1}{-2.5em}{0em}

\autoref{fig:ResultPerfBW} shows that full-stack optimization consistently yields the best performance per BW/NPU across both systems.
This highlights the importance of jointly optimizing workloads, collectives, and network components.
For System 1 and System 2, the full-stack optimization yielded performance improvements of 1.50--48.41$\times$ and 3.15--17.67$\times$, respectively, over single-stack baselines.

Among the single-stack variants:
\begin{itemize}[leftmargin=*]
    \item Collective-only optimization provided the least gain, as it only improves communication efficiency.
    \item Network-only optimization achieved more benefits by improving both communication and BW/NPU ratios.
    \item Workload-only optimization had the highest single-stack gains, as it can influence communication patterns and compute communication overlap.
\end{itemize}

\autoref{fig:ResultPerfCost} reports performance per dollar cost.
Here, full-stack optimization led to even greater improvements: 3.94--127.17$\times$ for System 1 and 3.40--38.73$\times$ for System 2, relative to single-stack baselines.
Interestingly, for System 2, network-only optimization outperformed workload-only optimization due to a significant impact on cost, despite smaller speedups.
\autoref{table:searched_parameters} summarizes the full-stack DSE results for System 2. 
Depending on the applied optimization constraint—BW/NPU or network dollar cost—the ML agent produces entirely different network configurations. 
These network choices, in turn, influence other design stacks, including workload parallelization strategies and selected collective algorithms. 
This highlights the tightly coupled nature of the ML system design space and reinforces the importance of performing full-stack optimizations.

In summary, full-stack system design consistently outperformed partial-stack approaches in both performance-per-BW/NPU and performance-per-cost metrics.
Across all experiments, full-stack optimization delivered at least 50--240\% improvement, underscoring the value of holistic system design made possible by \psa\ and its integration within \ours.

\begin{footnotesize}
\begin{table}[t]
\centering

\caption{System configurations used for evaluation.}
\label{table:searched_parameters}
\vspace{-1em}
\resizebox{\linewidth}{!}{

\begin{tabular}{lll}
\toprule
\textbf{Target} & \textbf{Perf per BW/NPU} & \textbf{Perf per Network Cost} \\
\midrule
\multicolumn{3}{l}{\textbf{Workload Knob}} \\
\midrule
\textbf{DP} & 64 & 128 \\
\textbf{PP} & 1 & 1 \\
\textbf{SP} & 4 & 4 \\
\textbf{Weight Sharded} & 1 & 1 \\
\midrule
\multicolumn{3}{l}{\textbf{Collective Knob}} \\
\midrule
\textbf{Scheduling Policy} & LIFO & LIFO \\
\textbf{Collective Algorithm} & {[}RI, RHD, DBT, DBT{]} & {[}RHD, RI, RI, RHD{]} \\
\textbf{Chunks per Collective} & 4 & 16 \\
\textbf{Multi-dim Collective} & Baseline & Baseline \\
\midrule
\multicolumn{3}{l}{\textbf{Network Knob}} \\
\midrule
\textbf{Topology} & {[}RI, RI, RI, SW{]} & {[}SW, FC, FC, FC{]} \\
\textbf{NPUs per Dim} & {[}4, 4, 4, 16{]} & {[}4, 8, 4, 8{]} \\
\textbf{Bandwidth per Dim} & {[}50, 50, 50, 50{]} & {[}50, 50, 50, 50{]} \\
\bottomrule
\end{tabular}
}

\end{table}
\end{footnotesize}

\subsection{Scalability Analysis}\label{section:Results:Scalability}

Next, we demonstrate the scalability of \psa\ and \ours\ for larger systems and workloads.
In this section, we leverage system 3 as the target baseline, which consists of 2,048~NPUs and is 4$\times$ and 2$\times$ larger in number of NPUs than system 1 and 2 used in~\autoref{section:Results:FullStack}, respectively.
For this larger target system, we use two distinct workloads with different scales: one set of experiments using a relatively small, \vitlarge\ model, and the other experiments targeting the larger \gptlarge.
Finally, we also execute DSEs with different global batch sizes, from 1,024 to 16,384 samples, to explore the scalability of \ours\ in terms of batch sizes (thereby increasing computation and communication requirements) as well.
Since workload-only optimization overall provided the best performance within the single-stack optimization target, thus we used workload-only optimization as our target baseline to compare full-stack optimizations.

\autoref{fig:ResultScalability} summarizes the experimental results for \ours\ scalability.
It's worth noting that:
\begin{itemize}[leftmargin=*]
    \item Regardless of the increased system size (2,048~NPUs), the full-stack optimization consistently showed great performance benefits over the workload-only optimization.
    \item Also, the performance benefit of the full-stack search was consistent regardless of the workload size: \vitlarge\ or \gptlarge.
    \item Finally, for every batch size, full-stack optimization always yielded better performance than workload-only optimization as well.
\end{itemize}

\insertFigure{ResultScalability}{
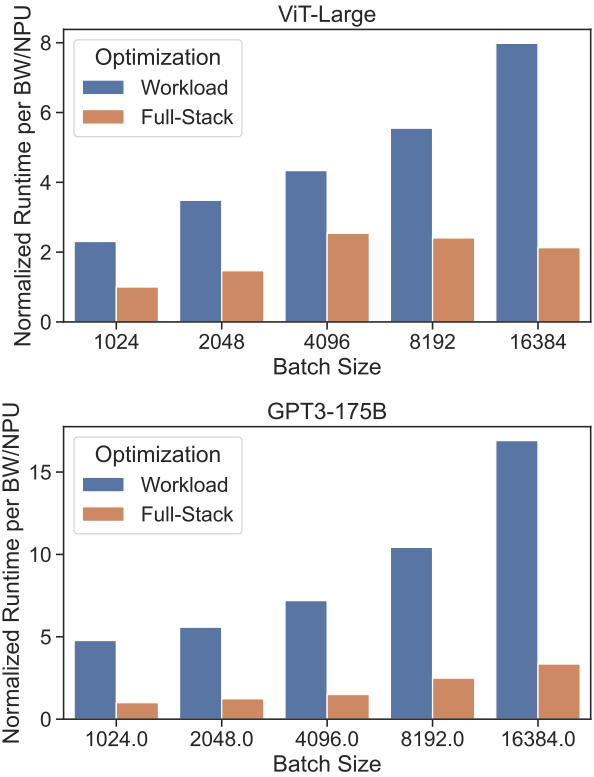
Workload-only vs. full-stack DSE result by \ours.
The target was system 3 with 2,048 NPUs.
Two workloads with different scales, \vitlarge\ and \gptlarge\ were used as the target model.
All results were normalized over the corresponding full-stack optimization result with 1,024 batches.
}{1}{-3em}{0em}

For \vitlarge, the performance benefit of the full-stack-optimized system over the workload-optimized cluster was 2.30$\times$, 2.38$\times$, 1.71$\times$, 2.31$\times$, 3.75$\times$ for global batch sizes 1,024--16,384.
Full-stack optimization provided at least 71\% performance improvement over the workload-only design.
Likely, \gptlarge\ workload provided 4.77$\times$, 4.52$\times$, 4.81$\times$, 4.19$\times$, and 5.05$\times$ over increasing global batch sizes, respectively.
The benefit for full-stack optimization was larger for \gptlarge\ than \vitlarge, showing at least 4.19$\times$ benefit compared to the 1.71$\times$.
Larger models, especially over large clusters, show higher resource demands in both compute and communication.
Therefore, judicious DSE and optimization are pivotal for bigger models over larger clusters.
Furthermore, it denotes more potentials that the full-stack co-design contains over a single-stack optimization, as exemplified by the experimental results in~\autoref{fig:ResultScalability}.
Therefore, we claim that the enabler for full-stack DSE, co-design, and optimization, such as the abstraction of \psa\ and \ours\ exploration capabilities, will get even more importance as the model and the cluster scales further.

\subsection{Co-Design Use Cases}

While~\autoref{section:Results:FullStack} and~\autoref{section:Results:Scalability} focused on full-stack optimization, the versatility of \psa\ enables DSE at arbitrary levels of the ML system stack. 
To demonstrate this flexibility, we conducted two additional co-design experiments, with results summarized in~\autoref{table:codesign}.

\niparagraph{Experiment 1} targets workload–network co-design while keeping collective algorithms fixed. 
This experiment jointly optimizes across all four workloads, allowing \ours\ to perform DSE across diverse training requirements. 
Interestingly, when optimizing for an ensemble of models, \ours\ consistently reduces memory footprint by increasing the TP size. 
To accommodate the enlarged TP groups, \ours\ allocates 64 NPUs along the first two dimensions—matching the TP group size exactly, and illustrating \ours's ability to co-optimize workload and network topology. 
Since the SP size is 8 and spans the final dimension, \ours\ selected the FC topology to avoid congestion, demonstrating intelligent adaptation to overlapping parallelism strategies.

\niparagraph{Experiment 2} explores collective–network co-design while fixing the workload parallelization strategy. 
This experiment targets inference for \gptlarge\ and includes two sub-scenarios: experiment 2.1 (Chat) and experiment 2.2 (question-answer).

\begin{footnotesize}
\begin{table}[t]
\centering
\newcommand{\myfix}{\textbf{\color{cyan}Fixed}}
\newcommand{\mysearch}{\textbf{\color{red}Searching}}
\newcommand{\mydse}{\textbf{\color{YellowOrange}Objective}}
\newcommand{\myderived}{\textbf{\color{Green}Derived}}
\newcommand{\myRed}[1]{\textcolor{red}{#1}}
\newcommand{\myOrange}[1]{\textcolor{YellowOrange}{#1}}
\newcommand{\myBlue}[1]{\textcolor{Green}{#1}}
\newcommand{\myPurple}[1]{\textcolor{cyan}{#1}}

\caption{\ours\ designed parameters for different use cases.}
\vspace{-1em}
*: \mysearch: Searched parameter; \myfix: Fixed value; \\
\vspace{0.5em}

\label{table:codesign}
\resizebox{\linewidth}{!}{

\begin{tabular}{l|cp{0.2em}@{}cp{0.2em}@{}cp{0.2em}@{}c}
\hline\hline

\textbf{} &
\textbf{Expr. 1} && \textbf{Expr. 2.1} && \textbf{Expr. 2.2}
\\ \hline
\textbf{Observations} & 
\textbf{Multi-Model} && \textbf{Chat Inference} && \textbf{QA Inference} 
\\ \hline

\multicolumn{1}{l}{\textbf{Network Knobs}}\\ \hline

Topology & 
\makecell[l]{ \myRed{[RI, FC, RI, FC]} } && 
\makecell[l]{ {\myRed{[RI, RI, RI, RI]}} } &&
\makecell[l]{ {\myRed{[RI, RI, RI, RI]}} } && 
\\ \hline

NPUs-count & 
\makecell[l]{ \myRed{[16, 4, 4, 4]} } && 
\makecell[l]{ {\myRed{[4, 16, 4, 4]}} } &&
\makecell[l]{ {\myRed{[16, 4, 4, 4]}} } && 
\\ \hline

Bandwidth per Link & 
\makecell[l]{ \myPurple{[50, 50, 50, 50]} } && 
\makecell[l]{ \myPurple{[50, 50, 50, 50]} } &&
\makecell[l]{ \myPurple{[50, 50, 50, 50]} } && 
\\ \hline

\multicolumn{1}{l}{\textbf{Collective Knobs}}\\ \hline

Scheduling Policy & 
\makecell[l]{ \myPurple{LIFO} } && 
\makecell[l]{ \myRed{LIFO} } &&
\makecell[l]{ \myRed{LIFO} } && 
\\ \hline

Chunks per Collective & 
\makecell[l]{ \myPurple{4} } && 
\makecell[l]{ \myRed{2} } && 
\makecell[l]{ \myRed{2} } &&
\\ \hline

Collective Algorithm & 
\makecell[l]{ 
    \myPurple{[RI, RI, RI, RI]} 
} && 
\makecell[l]{ 
    \myRed{[RHD, DBT, DBT, DI]} 
} &&
\makecell[l]{ 
    \myRed{[DI, DI, DBT, RHD]} 
} && 
\\ \hline

Multi-dim Collective & 
\makecell[l]{ \myPurple{Baseline} } && 
\makecell[l]{ \myRed{BlueConnect} } &&
\makecell[l]{ \myRed{Baseline} } &&
\\ \hline

\multicolumn{1}{l}{\textbf{Workload Knobs}}\\ \hline

Number of NPUs & 
\makecell[l]{ \myPurple{1024} } && 
\makecell[l]{ \myPurple{1024} } &&
\makecell[l]{ \myPurple{1024} } && 
\\ \hline

DP, PP, SP, TP & 
\makecell[l]{ \myRed{2, 1, 8, 64} } && 
\makecell[l]{ \myPurple{8, 4, 8, 4} } &&
\makecell[l]{ \myPurple{8, 4, 8, 4} } && 
\\ \hline

Weight Sharded & 
\makecell[l]{ \myRed{1} } && 
\makecell[l]{ \myPurple{1} } &&
\makecell[l]{ \myPurple{1} } && 
\\ \hline

\hline\hline
\end{tabular}

}
\undef{\myRed}
\undef{\myOrange}
\undef{\myBlue}
\undef{\myPurple}

\end{table}
\end{footnotesize}

As both use cases represent inference tasks, latency-optimized collectives are preferred over bandwidth-optimized ones due to the small message sizes during the decode phase. 
Accordingly, \ours\ consistently selected latency-efficient collective algorithms—Direct, RHD, and DBT—while avoiding Ring, which is typically bandwidth-optimized. 
Still, \ours\ opted to split collective operations into two chunks to enable pipelining during larger collectives encountered in the prefill phase of inference.

The results in~\autoref{table:codesign} underscore the generality of \ours. 
Thanks to the abstraction provided by \psa, \ours\ supports a range of co-design scenarios—including multi-model training and distinct inference workloads—without requiring any system-specific customization. 
\ours\ seamlessly adapts to partial-stack co-designs and diverse target objectives, showcasing its versatility in distributed ML system optimization.

\subsection{Comparing ML Agents}

\insertFigure{spiderplot_2x2}{
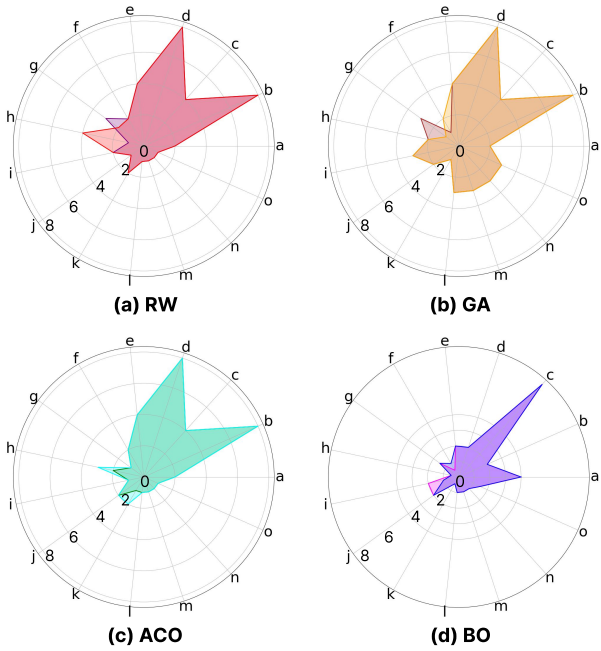
Differing configurations discovered within and across ML agents, all achieving equivalent optimal performance. Parameters are indexed. \textbf{a)}: Chunks per Collective; \textbf{b--e)}: 4D NPU count; \textbf{f)}: Scheduling Policy (1=FIFO, 2=LIFO); \textbf{g--j)}: 4D All-Reduce (1=RI, 2=DI, 3=RHD, 4=DBT); \textbf{k)}: Multi-dim Collective (1=Baseline, 2=BlueConnect); \textbf{l--o)}: 4D Topology (1=RI, 2=FC, 3=SW).
\label{fig:spider_plot}
}{1}{-2em}{-1em}

We now compare the effect of different ML agents when executing full-stack DSE in \ours{}.
As described in~\autoref{subsec:Subsec5-1}, \ours{} supports any ML and RL agents without modification, due to the modular abstraction provided by \psa{}.
This section evaluates \ours{} using four distinct ML agents, comparing their final design outputs and convergence behavior.

\autoref{fig:spider_plot} shows two best-performing system configurations for each agent in experiment 2 of~\autoref{table:codesign}, targeting \gptlarge\ training.
Although all agents converged to the same optimal reward, each discovered different design points.
Each agent identified multiple configurations yielding equivalent performance by preserving key parameters while exploring variations in others—highlighting the flexibility and redundancy in the design space.
This diversity provides system designers with a range of high-performing configurations, enabling trade-off considerations such as cost, complexity, or deployment constraints.

The variation arises from each agent’s unique exploration strategy, initialization behavior, and hyperparameter settings.
Notably, each configuration pair per agent in~\autoref{fig:spider_plot} reveals internal consistency in key performance-critical dimensions, with variance in less impactful parameters.

We further compare convergence behaviors in~\autoref{fig:convergence_2x2}, illustrating reward progression over 1,200 optimization steps.
The RW agent does not leverage history, resulting in a relatively flat reward curve—achieving high performance purely by chance.
In contrast, GA, BO, and ACO exhibit learning behaviors.
These agents progressively exploit high-reward regions while avoiding unproductive areas of the search space, as evidenced by upward trends followed by convergence.

All agents ultimately identified multiple optimal full-stack configurations, reaffirming the robustness of \ours{} across algorithmic choices.
The number of steps required to reach peak performance was:
RW — 652 steps; GA — 440 steps; ACO — 297 steps; BO — 680 steps.
These results suggest that more intelligent, history-aware agents can converge more quickly and avoid invalid configurations more effectively.

Crucially, \ours{} imposes no constraints on agent selection due to the abstraction provided by \psa{}.
This enables broad applicability to varied optimization goals, use cases, or user preferences.
As demonstrated here, \ours{} provides a powerful, agent-agnostic framework for full-stack system co-design at scale.

\insertFigure{convergence_2x2}{
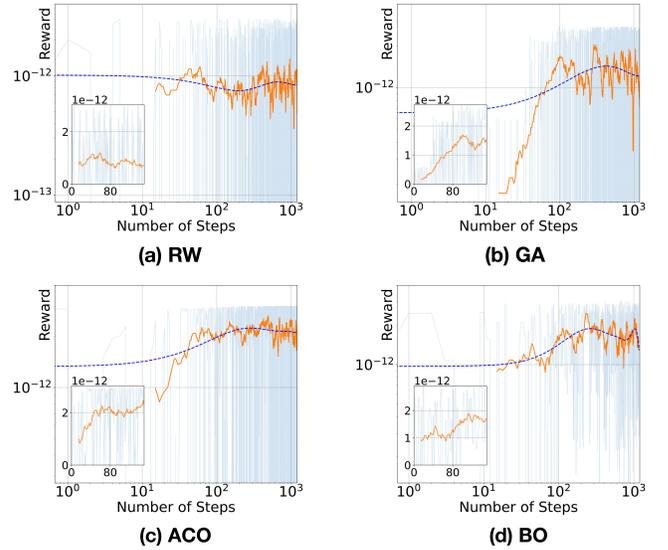
Reward vs. step convergence curves for each ML agent in experiment 4, from 0 to 1,200 steps.
\label{fig:convergence}
}{1}{-2em}{-1em}

\section{Related Work}\label{sec:RelatedWork}

Prior works~\cite{lu2017flexflow, lin2024uniap, Shah2023taccl, won2024tacos, won2024libra, zhao2024efficient, wang2023topoopt, kao2020gamma} have explored DSE for distributed ML, focusing on optimizing specific components of the system stack.
Workload-level optimization targets parallelization strategies given fixed system properties, as seen in FlexFlow~\cite{lu2017flexflow} (using Markov chain Monte Carlo) and UniAP~\cite{lin2024uniap} (using integer quadratic programming). 
Collective-level DSE focuses on identifying optimal collective algorithms for given network characteristics, exemplified by TACCL~\cite{Shah2023taccl} and TACOS~\cite{won2024tacos}.
Network-level optimization aims to improve collective performance by tuning bandwidth allocation~\cite{won2024libra} or synthesizing efficient topologies~\cite{zhao2024efficient}. 
TopoOpt~\cite{wang2023topoopt} further explores co-optimizing topology and parallelization using optical network reconfiguration. 
At the compute level, GAMMA~\cite{kao2020gamma} uses genetic algorithms~\cite{katoch2021ga} to optimize individual node performance but without considering any of the distributed ML contexts.

In contrast to these isolated approaches, we emphasize the need for full-stack co-design.
\ours{} enables holistic system optimization, uncovering bottlenecks across workload, system, network, and compute layers, and leveraging diverse search agents~\cite{katoch2021ga, dorigo1999aco, pearson1905rw, movckus1975bo} to identify non-obvious design points.
This approach allows us to integrate and analyze the impact of optimizations proposed in prior work within a unified framework, providing a better understanding of their effects on end-to-end performance.

\section{Conclusion}\label{sec:Conclusion}

We introduce \ours{}, a platform combining full-stack large-scale distributed ML system with agent-based search.
We propose the Parameter Set Architecture~(\psa) to abstract the interface between search agents and the system, enabling efficient exploration of design spaces with millions of parameters.
Our evaluation of transformer-based models demonstrates that \ours{} achieves 1.50--48.41$\times$ higher performance by optimizing the entire design stack.

\bibliographystyle{ACM-Reference-Format}
\bibliography{references}

\end{document}